# Arithmetic Operators over Finite Field GF($2^m$) for Error Correction Codes Application

Saeideh Nabipour
Faculty of Electrical and Computer Engineering
Mohaghegh Ardabili University
Ardabil, IRAN
Saeideh.nabipour@gmail.com

Masoume Gholizade
Faculty of Electrical and Computer Engineering
Semnan University
Semnan, IRAN
masoume_gholizade@semnan.ac.ir

**Abstract**: Galois field arithmetic circuits find application in a range of domains including error correction codes, communications, signal processing, and security engineering. This paper aims to elucidate the importance of error detection and correction techniques, while also scrutinizing the fundamental principles and wide array of techniques that can be employed. Additionally, a comprehensive understanding of the mathematical intricacies involved in BCH and Reed-Solomon codes requires extensive employment of GF($2^m$) arithmetic. Consequently, the primary contribution of this research is to critically examine the arithmetic operations performed over a finite field, which are essential for the successful implementation of BCH and Reed-Solomon codes. These operations encompass division, multiplication, exponentiation, multiplication inverses, addition, and subtraction.

**Keywords**: Finite field arithmetic, Galios field arithmetic, GF($2^m$) arithmetic operators, BCH Codes, Reed-Solomon Codes

1. **Introduction**

Information transmission between multiple components or from a single point to another, such as inside a computer network, may run into several challenges. Data may occasionally be intercepted by an intermediate computer system as a result of the existence of electromagnetic waves and a number of other reasons. In these situations, it is vital for the recipient to confirm the accuracy of received information. Error detection technique is an effective approach that can be applied in this situation. After detecting an error, a system with error detection capabilities can act in two different ways. The first option is to notify the sender of the error in the received information, prompting the sender to submit the prior information again. Automatic retransmission request, or ARQ, is the name of this technique. Use of error-correction methods is the second technique. A system with error correction capabilities can fix some errors without requiring the sender to send the message again. Choosing between these two approaches depends on a number of factors, some of which will be discussed, including the environment, line speed, and other factors. Consider a satellite transmission system as an illustration, in which data is sent from a satellite to a receiver on the ground. Due to the distance, it takes a data packet a lengthy time (about 250–300 milliseconds) to travel from the sender to the receiver, and a similar amount of time is needed for the receiver to identify a potential mistake in the packet. Sender should be informed of the receipt. When using the ARQ approach, the sender must store the delivered packets in its buffer until each one receives confirmation so that, if necessary, it can resend the earlier packets. If the transmitting speed is very fast, a very big buffer will be needed because of the lengthy delay between sending and receiving confirmation. FEC, or forward error correction, is more cost-effective in such circumstances and typically removes the need for retransmission. The theory of error detection and correction is a branch of engineering and mathematics that deals with reliable data transmission and storage. One of the characteristics of the 20th century has been the development and presentation of new storage and communication media. Simultaneously with the growth of information transmission and storage tools, information theory has been proposed. This theory was initially proposed by a person named Claude Elwood Shannon. He published his initial thoughts in 1948 in the article "Mathematical Theory of Communication". The issues that are investigated in the theory of information are: finding the best method for using the available communication systems and the best methods for separating information or desirable signals free from errors or unwanted information and improving communication channels for optimal communication. In general, the error correcting codes are divided into two categories, which are:

- Block codes: such as Hamming codes, Reed-Solomon codes, and BCH codes.
- Convolutional codes

When coding information, it is obvious to identify the difference between these two groups. In the first category, a block of input data is coded as a single unit or information package, resulting in the production of a code word that is a certain length. Both the input information packet and the output word code have constant lengths of k and n, respectively. The input information is not handled as a block in the second category of convolution codes; rather, it is fed into the encoder as a string of consecutive bits, and the coded information is produced as a string of bits. Information of various lengths may be encoded using a constant encoder circuit without modification. A block code system's encoder circuit, however, can only encode data that is up to a specific length, such as k bits. The check bits are consistently positioned adjacent to the information bits in this class of codes. The types of error correcting codes are displayed in Figure 1 [6].

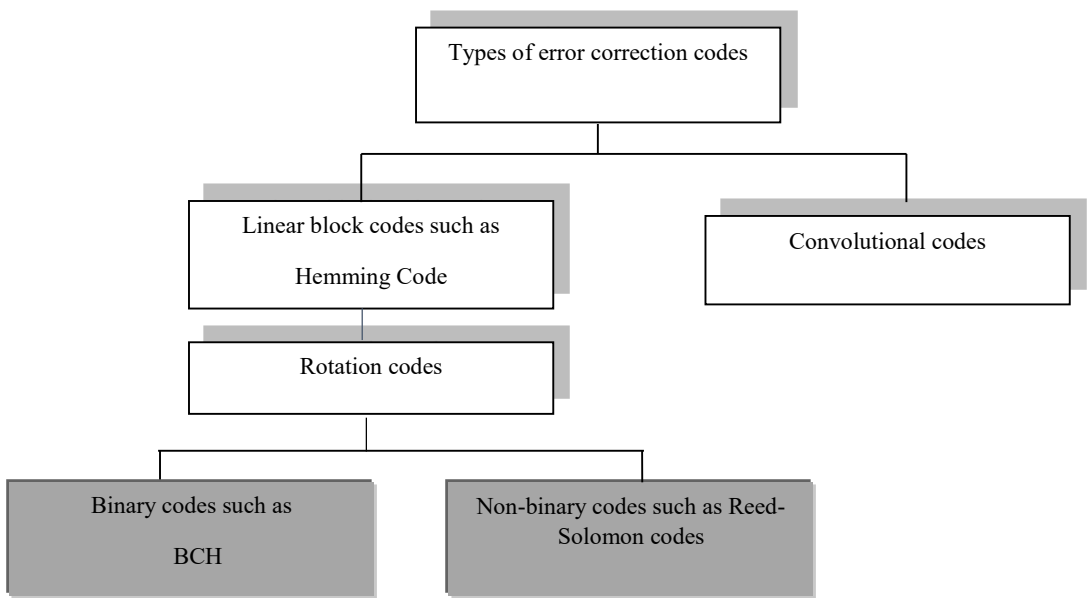

Figure 1: Different types of error correction codes [6]

In general, some additional data should be sent together with the primary information in any errors detection and correction system. A system that converts a k-bit message into a n-bit long code word is represented by pairs (n, k) and the ratio $R = \dfrac{k}{n}$ is known as the code rate, which is a way to quantify how much redundancy there is. if the R value is close to 1, it means that we are making better use of the available bandwidth and the message is not coded, and if the R value is close to zero, it means that we have more redundancy. Bit error rate (BER) vs signal-to-noise ratio (SNR) for both coded and un-coded systems is illustrated in Figure 2. Redundancy imposes a significant burden on the channel bandwidth and transmission power of the transmission resources; as a result, the excess quantity should be as little as possible. Coding rate and coding efficiency are two opposing factors. Although the error correcting capability has been improved, the coding rate falls as the message's redundancy increases. The performance of errors correction should be maximized while maintaining a coding rate close to 1.

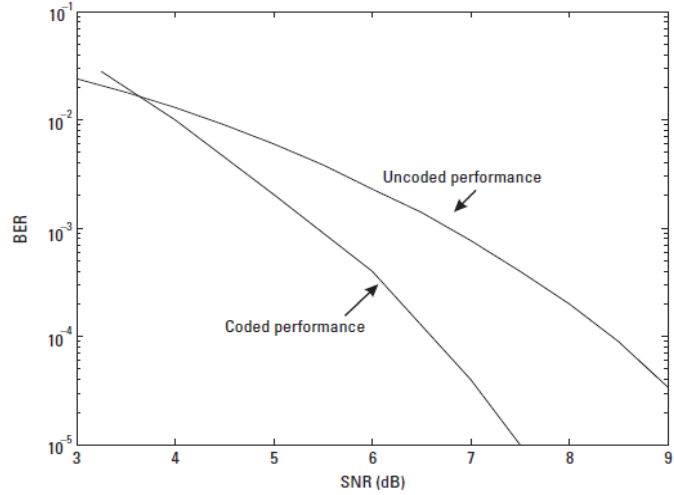

Figure 2: BER performance of coded systems

In this paper, we will explain and review the basic concepts and definitions in coding theory. In addition, explaining the mathematical details of BCH and Reed-Solomon codes requires the comprehensive use of modern algebra, so in the rest of this paper, we will review the introductory concepts of algebra and introduce the finite field arithmetic. Also, considering that the main operators needed to implement the BCH and Reed-Solomon codes are: addition, multiplication and division, in this paper we will examine these main components.

## 2. Basic Definitions

To continue the discussion about the quality of the codes, it is necessary to provide some definitions:

- **Hamming Distance:** The disparity between two code words is determined by the quantity of bits in which they diverge. For example, the following two words have a distance of 2, because they differ in 2 bits:

$$000$$

$$011$$

- **Minimum Distance:** the value of $d_{min}$ represents the smallest possible distance between valid code words within a given system; that's mean,

$$d_{min}(C) = \min\{d(x,y) \mid x, y \in C, x \neq y\}.$$

For example, in the system of the previous example (triple repetition), $d_{min}$ is the distance between two words 000 and 111, that is $d_{min} = 3$. In general, for a system to be able to detect t errors, we must have $d_{min} \geq t+1$ : and for a system to be able to correct t errors, we must have $d_{min} \geq 2t+1$. In a system with specifications (n, k), the maximum value $d_{min}$ can be $n-k+1$. In other words, a system with a certain $d_{min}$ value can detect the maximum error $d_{min}-1$ and correct the maximum error $\frac{d_{min}-1}{2}$. Therefore, the goal in designing a system is to raise R and increase the power of error detection and correction. To increase the power of detection and error correction, we must increase the value of n-k, i.e. redundancy, and to increase R, we must increase the ratio. In the previously mentioned triple repetition system, $k=1$ and $n=3$; So we will have:

$$\text{Rate Code: } R = \frac{1}{3}$$

$$\text{Minimum distance: } d_{\min} = n - k + 1 = 3$$

$$\text{Error detection capability: } d_{\min} - 1 = 2$$

$$\text{Error correction feature: } \frac{d_{\min} - 1}{2} = 1$$

- **Asymmetric Codes**: their ability to detect errors is limited to single-bit errors (0 to 1 or 1 to 0). This type of code is useful in programs that only expect an error to occur.
- **Non-Separable or Non-Systematic Codes:** Codes in which the check bits embedded in the code word cannot be processed simultaneously with information are called non-separable or non-systematic codes.
- **Linear Block Codes:** in this method, the set of both code words forms a new code word in the form of a block of bits in the source. Converting the original binary string to a coded string by block-by-block method, adding r additional bits to each block with n bits of information, forms the block code generation steps.
- **Rotating Codes:** they form a set of linear block codes, which generate a new code word for each type of rotation.
- **Forward Error Correction Codes:** in this approach, supplementary data is transmitted to the recipient in conjunction with the primary data and serves the purpose of identifying and rectifying any errors that occur during the recipient's processing. Consequently, there is no necessity to retransmit the aforementioned data.
- **Backward Error Correction Codes:** In this method, redundancy can only help the receiver to detect the presence of an error, and the information with the error must be sent again.

In channels without memory, noise affects every symbol sent without dependency. For example, in a symmetric binary channel, the probability P is assigned to each transmitted bit indicating both incorrect and correct received data, which is independent of other transmitted bits. Therefore, the occurrence of errors in the received data is random, resulting in the classification of channels lacking memory as channels with random errors. A prime illustration of these channels can be found in satellite channels and deep space channels. In actuality, the majority of direct line transfers exhibit random errors. Codes that are incorporated to rectify random errors are referred to as random error correction codes. A simplified representation of a memory channel can be observed in Figure 3. This model encompasses two modes: a favorable mode, in which errors transpire in a non-repetitive manner during each transfer, and an unfavorable mode, in which errors possess the highest likelihood in every transfer. In the majority of instances, the channel is in a favorable condition. However, when the channel transmission parameters are altered, such as in the scenario of deep fading due to multi-pathing, the occurrence of errors during transmission takes the form of clusters and bursts. This phenomenon arises from the increased likelihood of transitioning to an unfavorable state. Channels that possess a memory component are referred to as channels with burst or burst errors. An example of a channel that exhibits scattering errors is a radio channel, which experiences signal fading during multi-path transmission. Fading can also manifest in the transmission wire or cable due to switching noise or cross talk. The codes used to correct these types of errors are known as burst error correcting codes. Finally, some channels have a combination of both random error and dispersion and are called composite channels. The correcting codes for this type of errors are also called the random-burst error correcting codes [14].

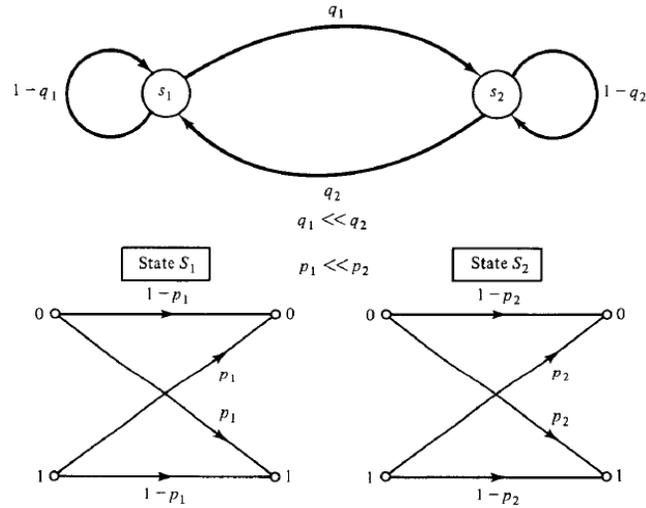

Figure 3: A simple model of a memory channel [8]

## 3. Fundamentals of Algebra

### 3-1. Groups

More specifically, a group can be described as an algebraic structure on an unbounded collection that demonstrates closure with respect to a binary operation and also participates in the relation associated with that operation. Also, the presence of the same element and the opposite element in this structure is mandatory. According to this definition: The following requirements must be met in order to classify (G, $\times$) as a group, if the group G is infinite and dual operation is on G:

1. For each $a, b \in G$, $a \times b \in G$ (G is closed to action $\times$)
2. For each $a, b, c \in G$, $a \times (b \times c) = (a \times b) \times c$ (participation feature)
3. For each $a \in G$, there exists an $e \in G$ such that $a \times e = e \times a = a$ (existence of the same element)
4. For each $a \in G$, there exists a, $b \in G$ such that $a \times b = b \times a = e$ (existence of inverse element)

In this case, G together with the dual operation is called group and we represent them with (G, $\times$). A group in mathematics is a collection that may be subjected to binary functions like multiplying, adding, etc. A collection of numbers is an illustration of a group over the addition function.

### 3-2. Field

A field in mathematical subjects is a structure based on algebra where the four functions of adding, subtracting, multiplying, and dividing are specified (aside from division by zero), and where both the functions of adding and dividing have commutative features.

A field F is a set of elements with two operations + (addition) and . (multiplication) which applies to the following feature:

- F is closed over + and . , that is, for a and b, a + b and a.b belong to F.

For each a, b and c in F, there are following features:

- Commutative feature: a + b = b + a, a.b = b.a

- Associative feature: (a + b) + c = a + (b + c)  a. (b.c) = (a.b). c

- Distributive feature: a. (b + c) = a.b + a.c

Furthermore, the identity element 0 and 1 must exist in F, such that there are following conditions:

- For each a in F, a + 0 = a.

- For each a in F, a.1 = a.

- For each a belonging to F, the inverse element of multiplication $a^{-1}$ exists in F such that $a.a^{-1} = 1$

The identical elements of the adding and multiplying operations are 0 and 1, respectively. The order of the field refers to a field's overall number of items. Keep in mind that the binary field only has two types of elements: 0 and 1. Coding theory, which we shall discuss in the following part, makes extensive use of this topic. The set of real numbers and the set of complex numbers are two instances of infinite fields. Because, for instance, the number 2 does not have a multiplicative inverse in this set, the set of integers is not a field.

**3-3. Galios Finite Field**

It is required to apply some algebraic concepts to the error correcting codes in order to make their usage and analysis more straightforward. It is extremely helpful to have an alphabet that allows for unlimited adding, subtracting, multiplying, and dividing. A finite field is a field with a finite amount of elements; this quantity is referred to as the field's order. French mathematician Evariste Galiosis established the fundamental conclusion regarding the finite field, which is shown in Theorem 1.2 below.

**Theorem 1.** if q is an exponent of a prime number (q = $p^m$ such that p is a prime value and m is a positive integer), then there is a field of degree q. Furthermore, there is only one field of that order, if q is an exponent of a prime integer.

A finite field, also known as a Galios field, is a field that has a limited number of members. The sign GF(q) is used to represent the Galios field of order q. Coding theory frequently makes utilization of this kind of field. A finite field is referred to be a primitive field if its order is the prime integer p. Of course, the elementary field with p=2 is the binary field. There exists a minimal positive integer for each Galios field GF(q) as follow:

$$\underbrace{1+1+...+1}_{\lambda 1's} = 0$$

The unique feature of the field is the integer value λ. The number 2 is a feature of the Galios field. because $1 \oplus 1 = 0$ P is a feature of the GF(p) initial field. In overall, a prime number must be an attribute of every Galios field. The expanded Galios field GF(p), which is represented as GF(pm) and is an expanded field, is the Galios field of the order $q = p^m$ that p is a prime number and m is a positive integer. The fundamental difference between a finite field and an unbounded field is that the bounded field operators have a hidden MOD (integer remainder) operator. For example, only the numbers 0 and 1 are significant over $GF(2)$ and 1 + 1 is equal to 0, because 2 mod 2 is zero. In fact, the operation in a limited field always produces a result in the field. The fact that the Galios field always contains at least one element whose powers constitute the set of all non-zero elements in the field—this element is referred to as the main element of the finite field—is a key characteristic of the field. If it is the field's principal element, the following results will be achieved for all of the field's non-zero elements:

$$\alpha^0(=1), \alpha^1, \alpha^2, \alpha^3, ..., \alpha^{q-2} \tag{1}$$

Equation (1) shows that a field can be made completely by its primary element and its powers. In other words, GF($2^m$) includes all combinations of m bits, which we represent as different powers of α, which is the primary element of GF. Thus, GF($2^m$) has $2^m$ elements. The power representation of the field elements is the exponential representation of the field elements according to equation (1). The following equation, where i is an integer, represents the element's *nth* exponent, if there is any random non-zero element in the finite field GF(q):

$$\beta^n = (\alpha^i)^n = \alpha^{i.n}$$

$i \cdot n$ must be a multiple of $q-1$ in order for::

$$\beta^n = \alpha^{i.n} = \alpha^{q-1} \cdot \alpha^{q-1} ... \alpha^{q-1} = 1$$

### 3-4. Galios Field and Polynomials

The polynomial representation of the elements is another way to express the power of the finite field's components. In this part, we look at the field GF($p^m$)'s polynomial representation. A one-variable polynomial with coefficients in the initial Galios field GF(p) is said to be over the field of degree n.

$$f(x) = f_0 + f_1 X + f_2 X^2 + f_3 X^3 + ... + f_n X^n \tag{2}$$

Each of the p components of the field GF(p) are the coefficients of the polynomials. The only difference between a polynomial over the field GF(p) and a typical polynomial is that the coefficients of the polynomial are derived using module 2 methods. The following equation is a beneficial property of polynomials defined over the field:

$$f^2(X) = f(X^2) \tag{3}$$

Equation (3) is obtained as follows:

$$f^2(X) = (f_0 + f_1 X + f_2 X^2 + ... + f_n X^n)^2$$
$$= [f_0 + (f_1 X + f_2 X^2 + f_3 X^3 + ... + f_n X^n)]^2$$
$$= f_0^2 + f_0(f_1 X + f_2 X^2 + f_3 X^3 + ... + f_n X^n) + f_0(f_1 X + f_2 X^2 + f_3 X^3 + ... + f_n X^n)$$
$$+ (f_1 X + f_2 X^2 + f_3 X^3 + ... + f_n X^n)^2 = f_0^2 + (f_1 X + f_2 X^2 + f_3 X^3 + ... + f_n X^n)$$

Therefore, solving the above equation will lead to the following equation:

$$f^2(X) = f_0^2 + (f_1 X)^2 + (f_2 X^2)^2 + ... + (f_n X^n)^2$$

Considering $f_i = 0,1$ and $f_i^2 = f_i$, the above equation is simplified as equation (4):

$$f^{2^i}(X) = f(X^{2^i}) \tag{4}$$

Equation (4)'s reasoning is that if $\beta$ is the answer of a polynomial over the Galios field, afterwards $\beta^{2^i}$ is also answer of that polynomial ($f(\beta^{2^i}) = f^{2^i}(\beta) = 0$); Therefore, existence $\beta$ as a polynomial root also refers to existence of $\beta^2, \beta^{2^2}, \beta^{2^3}, \ldots$ as the roots of that polynomial.

**Definition of Irreducible Polynomial:** A polynomial $\mathcal{H}(X)$ is irreducible if it cannot be divided by any polynomial of degree lower than m over the field GF(p) and has degree m or more. A polynomial, for instance, $X^3 + X + 1$ is an irreducible polynomial over $GF(2)$; However, polynomial $X^3 + X$ is different because $(X^3 + X)/X = X^2 + 1$

**Theorem 2.** A divisor of the polynomial $X^{p^m-1} + 1$ is an irreducible polynomial of order m under. For example, we can check that a polynomial $X^3 + X + 1$ is a divisor of a polynomial $X^{2^3-1} + 1 = X^7 + 1$:

```
                    x⁴ + x² + x + 1
          ┌─────────────────────────
x³ + x + 1│ x⁷ + 1
          │ x⁷ + x⁵ + x⁴
          │ ─────────────
          │ x⁵ + x⁴ + 1
          │ x⁵ + x³ + x²
          │ ─────────────
          │ x⁴ + x³ + x² + 1
          │ x⁴ + x² + x
          │ ─────────────
          │ x³ + x + 1
          │ x³ + x + 1
          │ ─────────────
          │ 0
```

**Definition of Prime Polynomial $\varphi(X)$:** If a degree m irreducible polynomial $\varphi(X)$ is divided by $X^n + 1$ for positive integer n ($(n = 2^m - 1)$), then it is a prime polynomial. For instance, a polynomial $\varphi(X) = X^4 + X + 1$ is divisible by $X^{15} + 1$, while this polynomial will not be divisible by $X^n + 1$ for $1 \leq n \prec 15$; Therefore, the polynomial $\varphi(X) = X^4 + X + 1$ is a prime polynomial. In general, every Galios field $GF(2^m)$ has a prime polynomial of degree m, whose root is α. Recognizing the prime polynomial is not an easy task, however, a list of prime polynomials for different m is given in Table 1. It is important to note that there may be more than one prime polynomial per m; But the polynomial with the least number of sentences is considered as the prime polynomial. Prime polynomial plays a central role in coding theory. By defining the prime polynomial over the field, we can show all the elements of the field using polynomials.

Table 1: Prime polynomials over different m of the Galios field [6]

| m | | m | |
|---|---|---|---|
| 3 | $1+X+X^3$ | 14 | $1+X+X^6+X^{10}+X^{14}$ |
| 4 | $1+X+X^4$ | 15 | $1+X+X^{15}$ |
| 5 | $1+X^2+X^5$ | 16 | $1+X+X^3+X^{12}+X^{16}$ |
| 6 | $1+X+X^6$ | 17 | $1+X^3+X^{17}$ |
| 7 | $1+X^7+X^3$ | 18 | $1+X^7+X^{18}$ |
| 8 | $1+X^2+X^3+X^4+X^8$ | 19 | $1+X+X^2+X^5+X^{19}$ |
| 9 | $1+X^4+X^9$ | 20 | $1+X^3+X^{20}$ |
| 10 | $1+X^3+X^{10}$ | 21 | $1+X^2+X^{21}$ |
| 11 | $1+X^2+X^{11}$ | 22 | $1+X+X^{22}$ |
| 12 | $1+X+X^4+X^6+X^{12}$ | 23 | $1+X^5+X^{23}$ |
| 13 | $1+X+X^3+X^4+X^{13}$ | 24 | $1+X+X^2+X^7+X^{24}$ |

If $\varphi(X)$ is the prime polynomial of GF($p^m$) and i is an integer, then $X^i$ is defined as follows:

$$X^i = q(X)\varphi(X) + r(X) \tag{5}$$

That $q(X)$ and $\varphi(X)$ are quotient and remainders, respectively. If $\alpha$ is a root $\varphi(X)$ ($\alpha$ the initial element of the field is finite), by substituting $X = \alpha$ in equation (5), we will have:

$$\alpha^i = q(\alpha)\varphi(\alpha) + r(\alpha) \tag{6}$$

Given that $\varphi(\alpha) = 0$, therefore:

$$\alpha^i = r(\alpha) = r_0 + r_1\alpha + r_2\alpha^2 + ... + r_{m-1}\alpha^{m-1} \tag{7}$$

Equation (7) states that any arbitrary element $\alpha^i$ can be represented as a polynomial $r(\alpha)$, which is obtained as equation (8):

$$\alpha^i \Rightarrow X^i \bmod \varphi(X)|X = \alpha \tag{8}$$

Example: Consider Galios Field $GF(2^3)$. There are seven positive field components $\alpha^0(=1), \alpha, \alpha^2, ..., \alpha^6$. The degree 3 prime polynomial is $X^3 + X + 1$ as shown in Table 1-2. Using the prime polynomial, the polynomial representation of the field elements will be as follows:

$$\alpha^0 \Rightarrow 1$$
$$\alpha^1 \Rightarrow \alpha$$
$$\alpha^2 \Rightarrow \alpha^2$$

$$\alpha^3 \Rightarrow X^3 \bmod X^3 + X + 1|_{x=\alpha} = 1 + \alpha$$
$$\alpha^4 \Rightarrow \alpha \cdot \alpha^3 = \alpha(1+\alpha) = 2\alpha + \alpha^2$$
$$\alpha^5 \Rightarrow \alpha^1 \cdot \alpha^4 = \alpha(\alpha + \alpha^2) = \alpha^2 + \alpha^3 \Rightarrow 1 + \alpha + \alpha^2$$
$$\alpha^6 \Rightarrow \alpha^1 \cdot \alpha^5 = \alpha(\alpha^2 + \alpha^3) = \alpha^3 + \alpha^4 \Rightarrow 1 + (1+1)\alpha + \alpha^2 = 1 + \alpha$$

In addition to representing the elements of the Galios field as polynomials, these elements can also be represented as vectors. For example, the vector representation of the element $\alpha^5$ is as $\alpha^5 \Rightarrow 1+\alpha+\alpha^2 \Rightarrow (1111)$. Three different types of representation: exponential, polynomial and vector of field elements $GF(2^4)$ are shown in Table 2.

Table 2: Three different types of representation of field elements

| Exponential representation | Polynomial representation | Vector representation |
|---|---|---|
| 0 | – | 0 |
| $\alpha^0=1$ | 1 | 0001 |
| $\alpha$ | $\alpha$ | 0010 |
| $\alpha^2$ | $\alpha^2$ | 0100 |
| $\alpha^3$ | $\alpha^3$ | 1000 |
| $\alpha^4$ | $\alpha+1$ | 0011 |
| $\alpha^5$ | $\alpha^2+\alpha$ | 0110 |
| $\alpha^6$ | $\alpha^3+\alpha^2$ | 1100 |
| $\alpha^7$ | $\alpha^3+\alpha+1$ | 1011 |
| $\alpha^8$ | $\alpha^2+1$ | 0101 |
| $\alpha^9$ | $\alpha^3+\alpha$ | 1010 |
| $\alpha^{10}$ | $\alpha^2+\alpha+1$ | 0111 |
| $\alpha^{11}$ | $\alpha^3+\alpha^2+\alpha$ | 1110 |
| $\alpha^{12}$ | $\alpha^3+\alpha^2+\alpha+1$ | 1111 |
| $\alpha^{13}$ | $\alpha^3+\alpha^2+1$ | 1101 |
| $\alpha^{14}$ | $\alpha^3+1$ | 1001 |

In ordinary algebra, some polynomials with real coefficients have no real zero, but they may have complex roots. For example, $X^2+6X+25$ does not have any real roots, but it consists of two mixed conjugate roots, $-3+4i$ and $-3-4i$, where $i=\sqrt{-1}$. In this instance, it is possible that a polynomial with coefficients (0 and 1) over $GF(2)$ does not have a root in the field itself, but rather has a root in the extended field $GF(2^m)$. For example, $X^4+X^3+1$ is an irreducible polynomial under $GF(2)$; so, it does not consist an answer over $GF(2)$, but it consists of four answers in $GF(2^4)$. By substituting the values of $GF(2^4)$ based on table (2) in $X^4+X^3+1$, the elements $\alpha^7, \alpha^{11}, \alpha^{13}, \alpha^{14}$ will be the roots of this polynomial.

$$(\alpha^7)^4+(\alpha^7)^3+1=\alpha^{28}+\alpha^{21}+1=(1+\alpha^2+\alpha^3)+(\alpha^2+\alpha^3)+1=0$$

In fact, $\alpha^7$ is the root of $X^4+X^3+1$. Similarly, we can prove that $\alpha^{11}, \alpha^{13}, \alpha^{14}$ are also the roots of the equation. Since $\alpha^7, \alpha^{11}, \alpha^{13}, \alpha^{14}$ are the roots of the equation, $X^4+X^3+1$ must match $X^4+X^3+1$ in value. To prove this, it is enough to calculate their product:

$$(X+\alpha^7)(X+\alpha^{11})(X+\alpha^{13})(X+\alpha^{14})$$
$$=[X^2+(\alpha^7+\alpha^{11})X+\alpha^{18}][X^2+(\alpha^{13}+\alpha^{14})X+\alpha^{27}]$$
$$=(X^2+\alpha^8 X+\alpha^3 X)(X^2+\alpha^2 X+\alpha^{12})$$
$$=X^4+(\alpha^8+\alpha^2)X^3+(\alpha^{12}+\alpha^{10}+\alpha^3)X^2+(\alpha^{20}+\alpha^5)X+\alpha^{15}$$
$$=X^4+X^3+1$$

The answer to this question is given in the following theorem: Let's have a look at a polynomial $f(X)$ with coefficients in $GF(2)$. It is conceivable for a polynomial to have other field roots if β is a value of $GF(2^m)$ and a root of the polynomial $f(X)$. But whose field elements are these roots? The following theorem provides the solution to this query:

**Theorem 3.** $f(X)$ is a polynomial with coefficients over t $GF(2)$ and β is a value of $GF(2^m)$. If β is a root of $f(X)$, afterward for each $l \geq 0$, $f(X)$ also consists of $\beta^{2^l}$ as a root.

Proof: Equation (3) shows that we've got:

$$[f(X)]^{2^l} = f(X^{2^l})$$

When we substitute $\beta$ in the formula above, we get:

$$[f(\beta)]^{2^l} = f(\beta^{2^l})$$

Since $f(\beta) = 0$, $f(\beta^{2^l}) = 0$; so, $f(X)$ also consists of $\beta^{2^l}$ as a root..

$\beta^{2^l}$ is known as the conjugate of $\beta$. According to Theorem (3), all conjugates of $\beta$ that are elements of the field $GF(2^m)$ are roots of the polynomial $f(X)$ if $\beta$ is both an element of the field and a root of a polynomial $f(X)$ over $GF(2)$.

**Definition of Minimal Polynomial**: Suppose a polynomial over $GF(2)$ that is irreducible. A polynomial $\phi(X)$ is referred to a minimal polynomial of β. if it has the minimum degree over $GF(2^m)$ assuming that $\phi(\beta) = 0$. For instance, element zero's minimum polynomial over $GF(2^m)$ f is X, while element one's minimal polynomial is $X+1$.

The minimal polynomial of β can also be calculated from equation (9):

$$\phi(X) = \prod_{i=0}^{L-1}(X+\beta^{2^i}) \qquad (9)$$

The lowest number for which $\beta^{2^l} = \beta$ is L.

Example: Consider the Galios field $GF(2^4)$ whose elements are listed in Table (2). assuming that $\beta = \alpha^3$. The conjugate of $\beta$ is obtained as follow:

$$\beta^2 = \alpha^6, \quad \beta^{2^2} = \alpha^{12}, \quad \beta^{2^3} = \alpha^{24} = \alpha^9$$

The following steps are taken to find $\beta = \alpha^3$'s minimum polynomial:

$$\phi(X) = (X + \alpha^3)(X + \alpha^6)(X + \alpha^9)(X + \alpha^{12})$$

Thus, applying table (2) and multiplying the expression on the right side, gives us:

$$\phi(X) = [X^2 + (\alpha^3 + \alpha^6)X + \alpha^9][X^2 + (\alpha^{12} + \alpha^9)X + \alpha^{21}]$$

$$= (X^2 + \alpha^2 X + \alpha^9)(X^2 + \alpha^8 X + \alpha^6)$$
$$= X^4 + (\alpha^2 + \alpha^8)X^3 + (\alpha^6 + \alpha^{10} + \alpha^9)X^2 + (\alpha^{17} + \alpha^8)X + \alpha^{15}$$
$$= X^4 + X^3 + X^2 + X + 1$$

Table 3 shows the minimal polynomials of the elements over Galios field $GF(2^4)$.

Table 3: Minimal polynomials of Galios field elements

| Elements of Field | Minimal Polynomial |
|---|---|
| 0 | $X$ |
| 1 | $1 + X$ |
| $\alpha, \alpha^2, \alpha^4, \alpha^8$ | $1 + X + X^4$ |
| $\alpha^3, \alpha^6, \alpha^9, \alpha^{12}$ | $1 + X^2 + X^3 + X^4$ |
| $\alpha^5, \alpha^{10}$ | $1 + X + X^2$ |
| $\alpha^7, \alpha^{11}, \alpha^{13}, \alpha^{14}$ | $1 + X^3 + X^4$ |

There are different methods for the binary representation of Galios field elements, some of which are:

- Standard basis representation
- Normal basis representation
- Dual basis representation

**Standard Basis Representation:** In this method, each location of an m-bit vector represents one of the powers of α. In this way, the right-most place is related to $\alpha^0 = 1$ and the next place is related to $\alpha^1$ and in the same way the left-most place is related to $\alpha^{m-1}$. As a result, the Standard Basis representation of each $z \in GF(2^m)$ is as $z = a_{m-1}\alpha^{m-1} + a_{m-2}\alpha^{m-2} + \ldots + a_2\alpha^2 + a_1\alpha^1 + a_0$. As an illustration, to depict the elements of $GF(2^4)$, we act as follow:

$$0000 = 0$$
$$0001 = \alpha^0 = 1$$
$$0010 = \alpha^1$$
$$0100 = \alpha^2$$
$$1000 = \alpha^3$$

Other elements of the Galios field are built on the basis of its five primary elements. We can create all $GF(2^4)$ elements in following way:

$$0000 = 0$$
$$0001 = 1$$
$$0010 = \alpha$$
$$0100 = \alpha^2$$
$$1000 = \alpha^3$$
$$0011 = \alpha^4 = \alpha + 1$$
$$0110 = \alpha^5 = \alpha(\alpha+1) = \alpha^2 + \alpha$$
$$1100 = \alpha^6 = \alpha(\alpha^2 + \alpha) = \alpha^3 + \alpha^2$$
$$1011 = \alpha^7 = \alpha(\alpha^3 + \alpha^2) = \alpha^4 + \alpha^3 = \alpha^3 + \alpha + 1$$

**Dual Basis Representation**: according to the following two definitions:

**Definition 1**: Trace of $\beta \in GF(2^m)$ is as $Tr(\beta) = \sum_{k=0}^{m-1} (\beta)^{2^k}$.

**Definition 2**: $\mu_j$ and $\lambda_k$ are each other's dual, if:

$$Tr(\lambda_i \mu_j) = \begin{cases} 1 & if\ i = j \\ 0 & if\ i \neq j. \end{cases}$$

The dual basis representation of $\lambda_k$ is as $z = \sum_{k=0}^{m-1} z_k \lambda_k = \sum_{k=0}^{m-1} Tr(z\mu_k)\lambda_k$ and the elements of the Galios field are made based on four elements $\alpha^3, \alpha^2, \alpha, 1$.

**Normal Basis Representation:** its representation is as $z = a_{m-1}\alpha^{2m-1} + ... + a_2\alpha^2 + a_1\alpha^1 + a_0$ and members of the Galios field are made based on four elements $\alpha^3, \alpha^6, \alpha^{12}, \alpha^9$. Table 4 shows the three types of binary representation of Galios field over $GF(2^4)$.

Table 4: Three types of binary representation of members of the Galios field over $GF(2^4)$

| Power of α | Standard basis $1, \alpha, \alpha^2, \alpha^3$ | Dual basis $1, \alpha^3, \alpha^2, \alpha$ | Normal basis $\alpha^3, \alpha^6, \alpha^{12}, \alpha^9$ |
|---|---|---|---|
| - | - | 0000 | 0000 |
| 0 | 0000 | 1000 | 1111 |
| 1 | 1000 | 0001 | 1001 |
| 2 | 0100 | 0010 | 1100 |
| 3 | 0010 | 0100 | 1000 |
| 4 | 0001 | 1001 | 0110 |
| 5 | 1100 | 0011 | 0101 |
| 6 | 0110 | 0110 | 0100 |
| 7 | 0011 | 1101 | 1110 |
| 8 | 1101 | 1010 | 0011 |
| 9 | 1010 | 0101 | 0001 |
| 10 | 0101 | 1011 | 1010 |
| 11 | 1110 | 0111 | 1101 |
| 12 | 0111 | 0000 | 0000 |
| 13 | 1111 | 1000 | 1111 |
| 14 | 1011 | 0001 | 1001 |

What is usually used in hardware implementation of encoder and decoder circuits of error correction codes is the first method, which is also known as Canonical Basis and Polynomial Basis. Considering that multiplication is the most frequent calculation operation in the decoding process of BCH and Reed-Solomon error correction codes, Standard Basis multiplication is often preferred over Normal Basis and Dual Basis multiplication due to having the least hardware complexity in design.

### 4. Mathematical Operation over $GF(2^m)$

In digital communication and storage systems, data is represented in binary form. Considering that the function of error control codes is based on the binary field $GF(2)$ and $GF(2^m)$, this section deals with mathematical calculations in the field $GF(2^m)$.

### 4-1. Addition and Subtraction over Galios Field

Mathematical calculations in the Galios field, which has a finite number of values for each operand, are different from mathematical calculations of integers. Galios field calculations are done using module mathematics. The addition operation is the simplest operation in the Galios field, which is performed as a module 2 addition, or XOR operation between the corresponding bits in two elements of the Galios field. The subtraction operation in the field is exactly the same as the addition operation as follow:

$$u - v = u + v = u \oplus v, \qquad u, v \in GF(2)$$

As mentioned earlier, each element of the Galios field $GF(2^m)$ can be represented as a polynomial with coefficients over the GF(2) field; Therefore, the addition and subtraction of two elements over the Galios field $GF(2^m)$ is done as follow:

$$w(\alpha) = u(\alpha) + v(\alpha) = (u_0 + v_0) + (u_1 + v_1)\alpha + \ldots + (u_{m-1} + v_{m-1})\alpha^{m-1} \qquad (10)$$

That $u(\alpha) = u_0 + u_1\alpha + u_2\alpha^2 + \ldots + u_{m-1}\alpha^{m-1}$ and $v(\alpha) = v_0 + v_1\alpha + v_2\alpha^2 + \ldots + v_{m-1}\alpha^{m-1}$ are two elements of the Galios field $GF(2^m)$. The calculation of $u_i + v_i$ is done using the module 2 addition and the result will be as $w(\alpha) = w_0 + w_1\alpha + w_2\alpha^2 + \ldots + w_{m-1}\alpha^{m-1}$.

**Example**: suppose we want to add two elements $\alpha^7$ and $\alpha^{10}$ over the field $GF(2^4)$. According to table (2), the polynomial representation of these elements is as follow:

$$\alpha^7 \Rightarrow 1 + \alpha$$
$$\alpha^{10} \Rightarrow 1 + \alpha + \alpha^2$$

Therefore:

$$\alpha^7 + \alpha^{10} = (1+\alpha) + (1+\alpha+\alpha^2) = \alpha^2$$

Conducting addition and subtraction operators through the polynomial representation over the field $GF(2^m)$ is simple. According to the equation (10), the circuit of these two operations can be easily implemented using the XOR gate as shown in Figure 4.

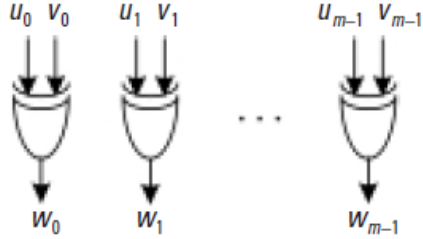

Figure 4: Galios field adder based on polynomials

## 4-2. Multiplication

The operation of multiplication over Galios field is performed by adding the multiplicand and multiplier powers together and using the module $(2^m - 1)$:

$$\alpha^i . \alpha^j = \alpha^{(i+j) \bmod 2^m - 1} \qquad (13)$$

which $\alpha^j$ and $\alpha^i$ are multiplicand and multiplier, respectively and the right side of the expression is equal to the product of the elements. When the operation of multiplication is performed using the polynomial representation, the product is equal to the multiplication of two polynomials module of the prime polynomial, in other words, the product must be reduced to the module of $\varphi(x)$ to be represented in m bits as follow:

$$w(\alpha) = u(x)v(x) \bmod \varphi(x) \big|_{x=\alpha} \qquad (14)$$

which $u(\alpha), v(\alpha) \in GF(2^m)$ are two multiplication factors, $w(\alpha)$ is the product and the $\varphi(x)$ is prime polynomial over the field $GF(2^m)$.

**Example**: If we use power representation to multiply two elements $\alpha^3$ and $\alpha^7$ over the field $GF(2^3)$, the result of the multiplication will be as follow:

$$\alpha^3 . \alpha^5 = \alpha^{(3+5) \bmod 2^3 - 1} = \alpha$$

On the other hand, if the polynomial representation of the elements is used, the multiplication is done as follow:

$$\alpha^3 . \alpha^5 = (1+X).(1+X+X^2) \bmod (1+X+X^3)\big|_{X=\alpha} = 1+X^3 \bmod (1+X+X^3)\big|_{X=\alpha} = \alpha$$

$1+X+X^3$ is the prime polynomial of the field $GF(2^3)$.

The operation of addition and multiplication is the most common operation in the Galios field. Unlike the adder, the multiplier circuit in Galios field is very complex, and due to the extensive use of error correction codes in the encoding and decoding process, it is crucial to apply considerable precision in their design. So far, many researches have been done to reduce the time overhead and the hardware complexity of the multiplier circuit. In general, the multipliers in the Galios field are divided into two categories:

- **Serial multipliers**: that receive the input bits one by one and are used in cases where high speed operation is not considered or in the case that very small circuits are needed.
- **Parallel multipliers**: that receive their two m-bit inputs simultaneously and have the structure of a combined circuit. These multipliers have very high speed and relatively large space.

In terms of input format, multipliers have different types, which include:

- Multipliers (Canonical/Polynomial Basis) Standard Basis
- Normal Basis multipliers
- Dual Basis multipliers
- Standard Basis Composite field multipliers

Each technique possesses distinctive attributes [1]. The choice of basis representation employed greatly influences the efficiency of the multiplication process in finite fields. Whereas dual basis multipliers necessitate extra modules for basis conversions, they typically demand a smaller chip area for implementation [14]. Division, multiplicative inversion, and exponentiation operations, conversely, are more suitable for NB base multiplication due to its preference for squaring in a finite field. RB bases eliminate the need for modulo reduction in addition to facilitating unrestricted squaring operations. The employment of RB bases, however, necessitates the inclusion of extra bits for the purpose of representing field elements. This, in turn, has the potential to lead to an augmentation in the intricacy of the hardware involved [15]. The polynomial basis multiplier, a well-established technique, offers an alternative that does not necessitate basis conversion. It presents an appealing option for hardware implementation owing to its consistency and simplicity. Additionally, PB (polynomial basis) multiplication can be categorized into parallel and serial calculations. Due to the fact that each clock produces all of the multiplication's output bits, parallel implementation yields great throughput. However, minimal space complexity is achieved through the utilization of bit-level serial computations. In the realm of bit-level serial multiplication methods, the decrease in area overhead is offset by a heightened computational delay, assessed by the quantity of clock cycles necessary for the creation of the m resultant bits. Subsequently, we will elaborate on the parallel PB multipliers.

**4-3. PB Parallel Multiplier based on Mastrovito Multiplication**

In (Zha, 2002), (Mas, 2006) and (Paa, 1994) various types of parallel multipliers have been discussed. According to the comparison made in (Paa, 1994), in fields smaller than 8 bits (i.e., $GF(2^m)$ with $m \prec 8$), the Standard Basis multiplier of the Mastrovito type is usually the best in terms of speed and the number of gates, or has small difference with the best technique. But in fields larger than 8 bits, the Composite Field multiplier, which is based on the Mastrovito multiplier, always gives the best results; Therefore, the best choice in building the internal circuits of the BCH decoder is the Mastrovito multiplier, which is discussed in detail in (Mas, 2006). Mastrovito proposed a different form of Standard Basis multiplication. Standard Basis multiplication consists of two steps: 1) multiplying the multiplicand polynomial by the multiplier 2) calculating the remainder of dividing the product by the prime polynomial of the Galios field.

If $\varphi(x)$ is the prime polynomial of the Galios field $GF(2^m)$ and the product of $a(x)$ and $b(x)$ is $C(x)$, then according to equation (14), the product of $a, b, c \in GF(2^m)$ is as follows:

$$a = a_0 + a_1\alpha + ... + a_{m-1}\alpha^{m-1}$$

$$b = b_0 + b_1\alpha + ... + b_{m-1}\alpha^{m-1}$$

$$c = c_0 + c_1\alpha + ... + c_{m-1}\alpha^m$$

$$C(x) = (a(x).b(x)) \bmod \varphi(x) \tag{15}$$

Mastrovito presented a type of parallel multiplication by using Z matrix multiplication to integrate the two steps mentioned in the Standard Basis multiplication. The multiplication operation will be done based on the Mastrovito multiplication according to the formula $c = Z.b$, where $c$ and $b$ are the coefficients of the product vector $C(x)$ and the multiplier $b(x)$, respectively, and the Z matrix is determined using the prime polynomial and the other multiplication factor $a(x)$. By rewriting equation (15), we will have the following form:

$$C(x) = [b_0 a(x) + b_1 x a(x) + ... + b_{m-1} x^{m-1} a(x)] \bmod \varphi(x) =$$
$$= [b_0 a(x) \bmod \varphi(x)] + [b_1 x a(x) \bmod \varphi(x)] + ... + [b_{m-1} x^{m-1} a(x) \bmod \varphi(x)]. \tag{16}$$

With definition of $Z_j(x)$, we have the following equation:

$$Z_j(x) = \sum_{i=0}^{m-1} z_{i,j} x^i = x^j a(x) \bmod \varphi(x) \qquad j = 0,1,...,m-1 \tag{17}$$

$$z_{i,j} \in GF(2)$$

As a result, using formula (17), the product is defined as equation (18):

$$C(x) = b_0 Z_0(x) + b_1 Z_1(x) + ... + b_{m-1} Z_{m-1}(x) = \sum_{j=0}^{m-1} b_j Z_j(x) \tag{18}$$

The product of the equation (18) is produced by using the Z matrix multiplication in the form of the equation (19):

$$C = \begin{bmatrix} c_0 \\ c_1 \\ ... \\ c_{m-1} \end{bmatrix} = \begin{bmatrix} z_{0,0} & z_{0,1} & \cdots & z_{0,m-1} \\ z_{1,0} & z_{1,1} & \cdots & z_{1,m-1} \\ \vdots & \vdots & & \vdots \\ z_{m-1,0} & z_{m-1,1} & & z_{m-1,m-1} \end{bmatrix} \bullet \begin{bmatrix} b_0 \\ b_1 \\ ... \\ b_{m-1} \end{bmatrix} = Z \cdot b \tag{19}$$

Z is called multiplication matrix. The Z matrix is a $m \times m$ binary matrix. Finding the multiplication matrix Z is the most difficult step in this type of multipliers. Each row of the matrix Z is a linear function of $a(x)$ and $\varphi(x)$, so each element of the Z matrix can be represented as:

$$C(x) = a(x)b(x) \bmod \varphi(x) = \sum_{i=0}^{m-1} c_i x^i \qquad (20)$$

$$C = \begin{bmatrix} c_0 \\ c_1 \\ \vdots \\ c_{m-1} \end{bmatrix} = Z \cdot b = \begin{bmatrix} f_{0,0} & f_{0,1} & \cdots & f_{0,m-1} \\ f_{1,0} & f_{1,1} & \cdots & f_{1,m-1} \\ \vdots & \vdots & & \vdots \\ f_{m-1,0} & f_{m-1,1} & & f_{m-1,m-1} \end{bmatrix} \cdot \begin{bmatrix} b_0 \\ b_1 \\ \vdots \\ b_{m-1} \end{bmatrix}$$

Functions $f_{i,j}$ are in the form of $f_{i,j}(a) = \sum_k a_k, k \in [0,1,\ldots,m-1]$. Mastrovito defined the multiplication matrix Z in the form of equation (21), its proof is given in (Mas, 2006).

$$Z = \begin{bmatrix} a_0 & a_{m-1} & a_{m-2} & \cdots & a_2 & a_1 \\ a_1 & a_0+a_{m-1} & a_{m-1}+a_{m-2} & \cdots & a_3+a_2 & a_2+a_1 \\ a_2 & a_1 & a_0+a_{m-1} & \cdots & a_4+a_3 & a_3+a_2 \\ \vdots & \vdots & \vdots & \vdots & \vdots & \vdots \\ a_{m-2} & a_{m-3} & a_{m-4} & \cdots & a_0+a_{m-1} & a_{m-1}+a_{m-2} \\ a_{m-1} & a_{m-2} & a_{m-3} & \cdots & a_1 & a_0+a_{m-1} \end{bmatrix} \qquad (21)$$

The parallel architecture of Mastrovito multiplication is shown in Figure 5. This architecture includes two blocks named f-network and IP-network. The f-network block produces the elements of the multiplication matrix Z, and the IP-network block implements the matrix-vector product of the equation (19) and includes m partial multiplication that calculates the following values:

$$c_i = [z_{i,0}, z_{i,1}, \ldots, z_{i,m-1}][b_0, b_1, \ldots, b_{m-1}]^T, \qquad 0 \leq i \leq m-1. \qquad (22)$$

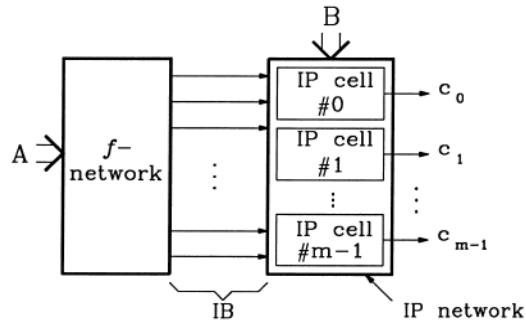

(a)

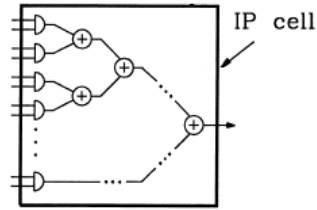

(b)

Figure 5: (a). Parallel architecture of Mastrovito multiplication over the field $GF(2^m)$ (b). IP block to perform partial multiplication

For example, consider the multiplication of two elements $A(x) = a_0 + a_1x + a_2x^2 + a_3x^3$ and $B(x) = b_0 + b_1x + b_2x^2 + b_3x^3$ over the field $GF(2^4)$ with prime polynomial $\varphi(x) = x^4 + x + 1$, according to the equation (26) and (28), the product will be as follow, which its parallel structure is based on the xor-tree in the IP block to calculate the product shown in Figure 6:

$$C = A \times B = \begin{bmatrix} a_0 & a_3 & a_2 & a_1 \\ a_1 & a_0+a_3 & a_3+a_2 & a_2+a_1 \\ a_2 & a_1 & a_0+a_3 & a_3+a_2 \\ a_3 & a_2 & a_1 & a_0+a_3 \end{bmatrix} \begin{bmatrix} b_0 \\ b_1 \\ b_2 \\ b_3 \end{bmatrix}$$

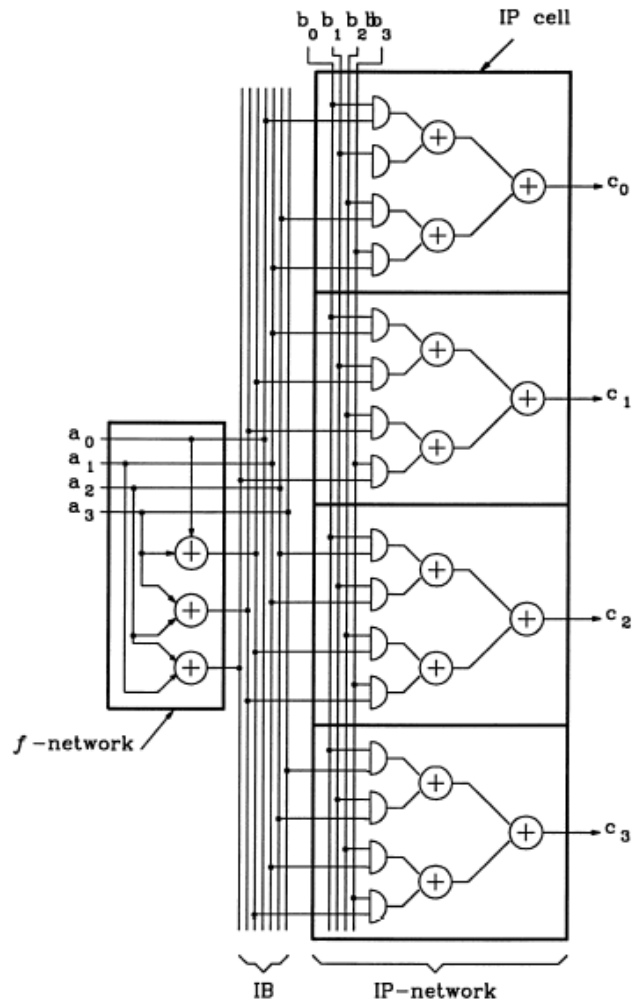

Figure 6: Parallel architecture of multiplication of two elements under the field

As a result, it can be said that Mastrovito bit-parallel polynomial basis multiplication is relatively difficult, but its main advantage compared to other parallel multipliers is the lowest delay. In some cases, it is necessary to always multiply by a constant number. For example, in the internal circuits of the encoder and decoder of the BCH and Reed-Solomon error correction codes, only multipliers are used that multiply their input by a constant number. In the Galios field, multiplication by a constant number can be designed much easier than a complete multiplier based on Mastrovito multiplication as follow:

$$\alpha^i b = \alpha^i (b_0 + b_1 \alpha + \ldots + b_{m-1} \alpha^{m-1}) \quad (23)$$
$$= b_0 \alpha^i + b_1 \alpha^{i+1} + \ldots + b_{m-1} \alpha^{i+m-1}$$

Considering that $\alpha^i$ is an m-bit element over the field $GF(2^m)$; as a result, the implementation of the constant multiplier using Mastrovito multiplication will lead to the production of the constant binary multiplication matrix Z according to equation (24):

$$\alpha^i b = Z.b = \begin{bmatrix} a_0^i & a_0^{i+1} & \ldots & a_0^{i+m-1} \\ a_1^i & a_1^{i+1} & \ldots & a_1^{i+m-1} \\ \vdots & \vdots & & \vdots \\ a_m^i & a_m^{i+1} & & a_m^{i+m-1} \end{bmatrix} \bullet \begin{bmatrix} b_0 \\ b_1 \\ \ldots \\ b_{m-1} \end{bmatrix} \quad (24)$$

The point worth mentioning in equation (2-24) is that the constant multiplier does not need any multiplication operation and its calculations are based on the module 2, or XOR gate and the number of XOR gate required to implement the constant multiplier over the Galios field is equal to:

$$\overline{\#XOR} = \frac{m^2}{2} - m. \quad (25)$$

In the following equations, you will see the equations related to constant multipliers. Note that here the operation + means XOR.

$$Output = Z = (z_3, z_2, z_1, z_0) \quad Input = A = (a_3, a_2, a_1, a_0)$$

$$Z = \alpha A = \begin{cases} z_0 = a_3 \\ z_1 = a_0 + a_3 \\ z_2 = a_1 \\ z_3 = a_2 \end{cases} \quad Z = \alpha^2 A = \begin{cases} z_0 = a_2 \\ z_1 = a_2 + a_3 \\ z_2 = a_0 + a_3 \\ z_3 = a_1 \end{cases} \quad Z = \alpha^3 A = \begin{cases} z_0 = a_1 \\ z_1 = a_1 + a_2 \\ z_2 = a_2 + a_3 \\ z_3 = a_1 + a_3 \end{cases}$$

$$Z = \alpha^4 A = \begin{cases} z_0 = a_0 + a_3 \\ z_1 = a_0 + a_1 + a_3 \\ z_2 = a_1 + a_2 \\ z_3 = a_2 + a_3 \end{cases} \quad Z = \alpha^5 A = \begin{cases} z_0 = a_2 + a_3 \\ z_1 = a_0 + a_2 \\ z_2 = a_0 + a_1 + a_3 \\ z_3 = a_1 + a_2 \end{cases} \quad Z = \alpha^6 A = \begin{cases} z_0 = a_1 + a_2 \\ z_1 = a_1 + a_3 \\ z_2 = a_0 + a_2 \\ z_3 = a_0 + a_1 + a_3 \end{cases}$$

$$Z = \alpha^7 A = \begin{cases} z_0 = a_0 + a_1 + a_3 \\ z_1 = a_0 + a_2 + a_3 \\ z_2 = a_1 + a_3 \\ z_3 = a_0 + a_2 \end{cases} \quad Z = \alpha^8 A = \begin{cases} z_0 = a_0 + a_2 \\ z_1 = a_1 + a_2 + a_3 \\ z_2 = a_0 + a_2 + a_3 \\ z_3 = a_1 + a_2 \end{cases} \quad Z = \alpha^9 A = \begin{cases} z_0 = a_1 + a_3 \\ z_1 = a_0 + a_1 + a_2 + a_3 \\ z_2 = a_1 + a_2 + a_3 \\ z_3 = a_0 + a_3 \end{cases}$$

$$Z = \alpha^{10} A = \begin{cases} z_0 = a_0 + a_2 + a_3 \\ z_1 = a_0 + a_1 + a_2 \\ z_2 = a_0 + a_1 + a_2 + a_3 \\ z_3 = a_1 + a_2 + a_3 \end{cases} \quad Z = \alpha^{11} A = \begin{cases} z_0 = a_1 + a_2 + a_3 \\ z_1 = a_0 + a_1 \\ z_2 = a_0 + a_1 + a_2 \\ z_3 = a_0 + a_1 + a_2 + a_3 \end{cases} \quad Z = \alpha^{12} A = \begin{cases} z_0 = a_0 + a_1 + a_2 + a_3 \\ z_1 = a_0 \\ z_2 = a_0 + a_1 \\ z_3 = a_0 + a_1 + a_2 \end{cases}$$

$$Z = \alpha^{13} A = \begin{cases} z_0 = a_0 + a_1 + a_2 \\ z_1 = a_3 \\ z_2 = a_0 \\ z_3 = a_0 + a_1 \end{cases} \quad Z = \alpha^{14} A = \begin{cases} z_0 = a_0 + a_1 \\ z_1 = a_2 \\ z_2 = a_3 \\ z_3 = a_0 \end{cases}$$

For example, to multiply a number by $\alpha^{13}$ over $GF(2^m)$, the circuit illustrated in Figure 7 can be used:

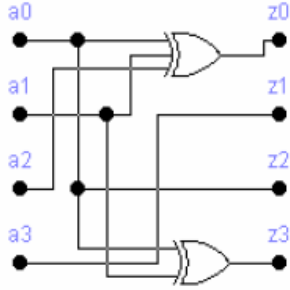

Figure 7: Constant multiplier circuit for $\alpha^{13}$

Motivated by the Mastrivoto multiplier, a novel modified algorithm has been proposed in [16] based on the method of serial interleaved multiplication. In order to achieve a more efficient utilization of area, a formulated algorithm has been suggested to eliminate the logical XOR (XNOR) gates utilizing a well-known logical relation. By employing this efficient logical relation, which takes advantage of the lower area and time complexities of the NAND gate compared to other gates such as AND or XOR/XNOR, hardware efficiency can be improved [16]. Their analysis demonstrates that the proposed multiplier achieves a low-area design in comparison to the majority of similar multipliers found in the literature, and it is comparable to the best existing area-efficient multiplier for m = 163. So, they proposed the equation (26) as follow:

$$P^K(x) = P^{(k-1)}(x) \oplus b_{m-k} A(x) \quad (26)$$

Since it is possible to discover that $(a \oplus b) = ((a \barwedge (a \barwedge b)) \barwedge ((a \barwedge b) \barwedge b))$ for each of the logic values a and b, we can change the logical XOR computation to an analogous circuit that just employs NAND as stated below, in this case, $\barwedge$ stands for logical NAND gate:

$$W^K(x) = W^{(k-1)}(x) \oplus (q_{m-k} \cdot P(x))$$
$$= W^{(k-1)}(x) \cdot \overline{q_{m-k} P(x)} + \overline{W^{(k-1)}(x)} \cdot q_{m-k} P(x)$$
$$= W^{(k-1)}(x) \cdot [\overline{W^{(k-1)}(x)} + \overline{q_{m-k} P(x)}] + [\overline{W^{(k-1)}(x)} + \overline{q_{m-k} P(x)}] \cdot q_{m-k} P(x)$$

$$= W^{(k-1)}(x)[\overline{\overline{W^{(k-1)}(x).q_{m-k}.P(x)}}].[\overline{\overline{W^{(k-1)}(x).q_{m-k}.P(x)}}.(q_{m-k}P(x)) \quad (27)$$

This equation shows that the logical NAND gate can be used to compute desired multiplication results rather of the logical XOR gate. Although compared to the conventional equation, this method can significantly improve hardware complexity which is desirable in constrained applications, such as smart cards, IoT devices, and implantable medical devices, the main disadvantage of their method is more critical path delay. The properties of this algorithm are discussed in more detail in [16]. As a summary, we list in Table 1 gate delays of PB multipliers based on the irreducible trinomial $u^n + u^k + 1$, where $2 < 2k < n$.

Table 5: Analyzing of time and hardware complexity for ($u^n + u^k + 1$)-based multipliers

| Multiplier | #AND | #NAND | #XOR | Critical path |
|---|---|---|---|---|
| PB Mastrovito [17, 18, 19, 20] | $2m^2 + 2m$ | 0 | $(n^2 - 1)$ | $T_A + \lceil \log_2 4n \rceil T_X$ |
| WDB [21] | $n^2$ | 0 | $(n^2 - 1)$ | $T_A + \lceil \log_2 4n \rceil T_X$ |
| PB mod reduction [22, 23] | $n^2$ | 0 | $(n^2 - 1)$ | $T_A + \lceil \log_2 (4n-4) \rceil T_X$ |
| PB Montgomery [24] | $n^2$ | 0 | $(n^2 - 1)$ | $\leq T_A + \lceil \log_2 (4n-8) \rceil T_X$ |
| WDB [25] | $n^2$ | 0 | $(n^2 - 1)$ | $T_A + \lceil \log_2 (2n+2k-2) \rceil T_X$ |
| PB Mastrovito [26] | $n^2$ | 0 | $(n^2 - 1)$ | $T_A + \lceil \log_2 (2n+2k-3) \rceil T_X$ |
| PB Mastrovito [27] | $n^2$ | 0 | $n^2 + (k^2 - 3k)/2$ | $T_A + \lceil \log_2 (2n+k-2) \rceil T_X$ |
| SPB Mastrovito [28] | $n^2$ | 0 | $n^2$ | $T_A + \lceil \log_2 2n \rceil T_X$ |
| SPB matrix–vector product [29] | $n^2$ | 0 | $3(n^2 - n)/2 - k(n - k)$ | $T_A + \lceil \log_2 (2n-k) \rceil T_X$ |
| PB Montgomery [30] | $n^2$ | 0 | $(n^2 - 1)$ | $T_A + \lceil \log_2 (2n-k) \rceil T_X$ |
| SPB binary XOR tree [31] | $n^2$ | 0 | $(n^2 - 1)$ | $T_A + \lceil \log_2 (2n-k) \rceil T_X$ |
| SPB Multiplier based on NAND [16] | $2m$ | $8m$ | 0 | $2T_A + 4T_N$ |

### 4-4. Divider

The most common method to calculate the quotient of the division $\frac{v}{u}$ over the field $GF(2^m)$ is to use the operation of multiplication between the dividend, v and the inverse of the divisor, $u^{-1}$; But the point that exists is how to calculate the inverse of an element over the Galios field. In general, a method for making a divider in the Galios field is to apply an inverter and a multiplier in sequence. So, we have to design an inverter. In the following, we will explain the construction of an inverter.

### 4-5. Inverse of Galios Field Elements

Suppose that the polynomial $\varphi(X)$ of degree m is an irreducible polynomial of the Galios field GF($2^m$); and β is a root of this polynomial $(\varphi(\beta) = 0)$. According to theorem (2-2), every irreducible polynomial is divisible by $X^{2^m-1}+1$; so we have:

$$X^{2^m-1}+1 = q(X)\varphi(X) \tag{28}$$

By replacing $X=\beta$, the following equation will be obtained:

$$\beta^{2^m-1}+1 = q(\beta).0 \Rightarrow \beta^{2^m-1}+1 = 0 \tag{29}$$

By adding 1 to both sides of the equation and using module 2, we have:

$$\beta^{2^m-1} = 1 \tag{30}$$

By multiplying both sides of equation (28-2) by $\beta^{-1}$, we have:

$$\beta^{2^m-2} = \beta^{-1} \tag{31}$$

This equation states that calculating the power $(2^m-2)$ of an element gives the inverse of that element. It is clear that the binary representation $(2^m-2)$ is in the form of 111...11 or $2^m-1 = 2^0 + 2^1 + 2^2 + ... + 2^{m-1}$ that can also be represented in the form of equation (2-30):

$$2^m - 2 = 2 + 2^2 + ... + 2^{m-1} \tag{32}$$

Therefore, the inverse of an element can be calculated as follow:

$$\beta^{-1} = \beta^{2^m-2} = \beta^2 \beta^{2^2} ... \beta^{2^{m-1}} \tag{33}$$

The implementation of equation (2-31) is shown in Figure 8, which includes a register, a multiplier and a squarer circuit. The register is initially loaded with 1. Considering that the circuit shifts in a rotational manner, the successive values of the register will be as follow:

$$1 \Rightarrow \beta^2 \Rightarrow \beta^6 \Rightarrow \beta^{14} \Rightarrow .... \Rightarrow \beta^{2^m-2} = \beta^{-1}$$

The final value of $\beta^{-1}$ is obtained after shift operation by $m-1$ times.

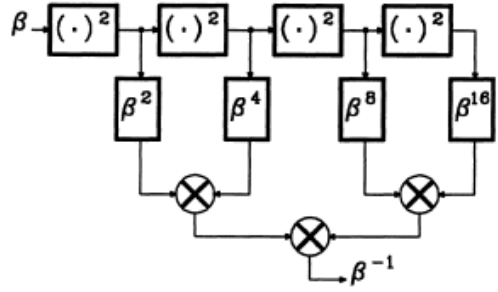

Figure 8: Inverse circuit using the power method

### 4-6. Exponent of two for Galios Field Elements

The exponent of two is one of the most used operators over the Galios field. Consider element $\alpha \in GF(2^m)$, whose polynomial representation is in the form of equation (2-32):

$$a = a_0 + a_1\alpha^1 + a_2\alpha^2 + ... + a_{m-1}\alpha^{m-1} \tag{34}$$

If $b \in GF(2^m)$ and $b = a^2$, according to the equation (2-4) ($f^2(x) = f(x^2)$), the expression of exponent of two of $a$ will be as follow:

$$b = a^2 = a_0 + a_1\alpha^2 + a_2\alpha^4 + ... + a_{m-1}\alpha^{2m-2} \tag{35}$$

In other words, the coefficients of $b$ can be obtained using the linear transformation of the coefficients of $a$ into the matrix form of equation (2-34), this matrix can be implemented using a number of XOR gates

$$b = a^2 = \begin{bmatrix} \alpha_0^0 & \alpha_0^2 & \alpha_0^4 & \cdots & \alpha_0^{2m+2} \\ \alpha_1^0 & \alpha_1^2 & \alpha_1^4 & \cdots & \alpha_1^{2m+2} \\ \vdots & \vdots & \vdots & \vdots & \vdots \\ \alpha_{m-1}^0 & \alpha_{m-1}^2 & \alpha_{m-1}^4 & \cdots & \alpha_{m-1}^{2m+2} \end{bmatrix} \begin{bmatrix} a_0 \\ a_1 \\ \vdots \\ a_{m-1} \end{bmatrix} \tag{36}$$

### 4-7. Linear Feedback Shift Register (LFSR)

A linear feedback shift register is a type of shift register that the input bit is a linear function of the previous state. In general, LFSRs are used in BCH and Reed-Solomon encoders, generating pseudo-random numbers, fast numerical counters, etc. This type of register is very suitable for hardware implementation and the systems made by them can be easily analyzed by linear algebra techniques. The structure of an LFSR consists of a number of D flip-flops and XOR gate. As shown in Figure 9, there are two types of linear feedback shift register: 1) internal feedback 2) external feedback.

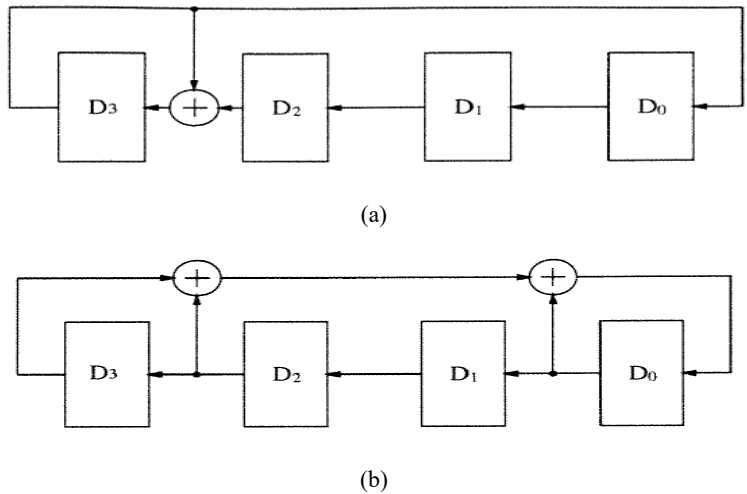

(a)

(b)

Figure 9: Two types of linear feedback shift register: (a). Internal feedback (b). External feedback

A polynomial can be easily represented using an LFSR, so that its coefficients are defined using the position of the XOR gates. The degree of the polynomial indicates the number of flip-flops and the feedback connections of the XOR gate to the flip-flops indicate the coefficients of the polynomial so that the presence of the connection indicates the coefficient 1 and its absence indicates the coefficient 0. For example, representation of $p(X) = X^3 + X + 1$ using LFSR is according to Figure 10. Considering that the polynomial $p(X)$ is of degree 3, we will have three D flip-flops and the location of the XOR gate is determined according to the coefficients of the polynomial as shown in Figure 10.

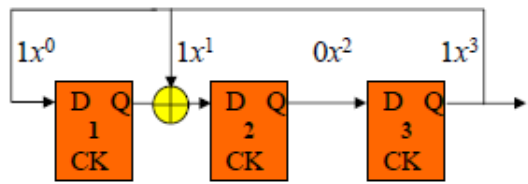

Figure 10: Polynomial representation using LFSR

One of the characteristics of LFSR is to perform multiplication and division operations in the encoder and decoder circuits of BCH and Reed-Solomon error correction codes. LFSR in Figure 11 and Table 6 shows the operation of dividing $P(x) = x^7 + x^6 + x^3 + x^2 + x$ by $G(x) = x^5 + x^3 + x^2 + 1$. P(x) = 11001110 enters the LFSR circuit from the most significant bit, and at the end of 8 clock pulses, the remainder of the division (R(x) = 01101) is stored in the flip-flops.

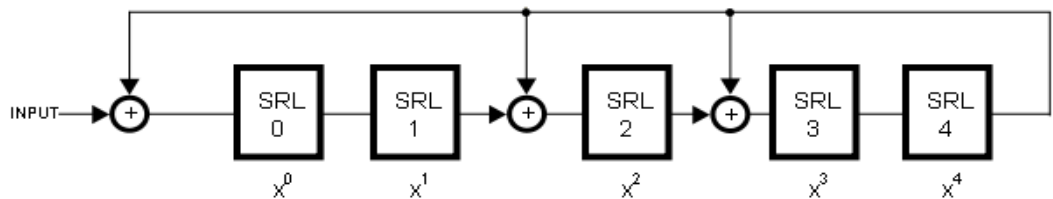

Figure 11. LFSR circuit for division $(x^7 + x^6 + x^3 + x^2 + x)/(x^5 + x^3 + x^2 + 1)$

Table. 6: Calculation of the remainder of division $(x^7 + x^6 + x^3 + x^2 + x)/(x^5 + x^3 + x^2 + 1)$ using LFSR

| Clock Cycle | Input | $X^0$ | $X^1$ | $X^2$ | $X^3$ | $X^4$ |
|---|---|---|---|---|---|---|
| 0 | – | 0 | 0 | 0 | 0 | 0 |
| 1 | 1 | 1 | 0 | 0 | 0 | 0 |
| 2 | 1 | 1 | 1 | 0 | 0 | 0 |
| 3 | 0 | 0 | 1 | 1 | 0 | 0 |
| 4 | 0 | 0 | 0 | 1 | 1 | 0 |
| 5 | 1 | 1 | 0 | 0 | 1 | 1 |
| 6 | 1 | 0 | 1 | 1 | 1 | 1 |
| 7 | 1 | 0 | 0 | 0 | 0 | 1 |
| 8 | 0 | 1 | 0 | 1 | 1 | 0 |

## 5. Conclusion

This paper discussed the algorithms and structures utilized in multiplicative groups, rings, and finite fields for the purpose of implementing BCH and Reed-Solomon error correction codes. The algorithms we presented have broad applicability, with a specific focus on the polynomial bases representation of elements within finite fields. We have provided a comprehensive overview of finite field arithmetic within error correction codes, encompassing all fundamental algorithms, structures, and components necessary to create efficient implementations of finite field operations. The fundamental arithmetic operations, namely addition, multiplication, division, and inversion, are performed within finite fields GF(q) where $q = p^k$, with p specifically being equal to 2.